\documentclass[prl,noshowpacs,preprintnumbers,amsmath,amssymb,twocolumn,superscriptaddress]{revtex4-1}

\usepackage{graphicx}
\usepackage{dcolumn}
\usepackage{bbold,bm}
\usepackage{enumerate}
\usepackage{bibunits}

\def\I{\text{I}}
\def\II{\text{II}}

\DeclareFontFamily{U}{rcjhbltx}{}
\DeclareFontShape{U}{rcjhbltx}{m}{n}{<->rcjhbltx}{}
\DeclareSymbolFont{hebrewletters}{U}{rcjhbltx}{m}{n}

\let\aleph\relax\let\beth\relax
\let\gimel\relax\let\daleth\relax

\DeclareMathSymbol{\aleph}{\mathord}{hebrewletters}{39}
\DeclareMathSymbol{\beth}{\mathord}{hebrewletters}{98}
\DeclareMathSymbol{\gimel}{\mathord}{hebrewletters}{103}
\DeclareMathSymbol{\daleth}{\mathord}{hebrewletters}{100}

\DeclareMathSymbol{\lamed}{\mathord}{hebrewletters}{108}
\DeclareMathSymbol{\mem}{\mathord}{hebrewletters}{109}
\DeclareMathSymbol{\ayin}{\mathord}{hebrewletters}{96}
\DeclareMathSymbol{\tsadi}{\mathord}{hebrewletters}{118}
\DeclareMathSymbol{\qof}{\mathord}{hebrewletters}{114}
\DeclareMathSymbol{\shin}{\mathord}{hebrewletters}{152}

\makeatother

\begin{document}
	
\begin{bibunit}	

\title{Non-Abelian fermion parity interferometry of Majorana bound states in a Fermi sea}

\author{D. Dahan}
\affiliation{Department of Physics, Ben-Gurion University of the Negev, Beer-Sheva
8410501, Israel}

\author{M. Tanhayi Ahari}
\affiliation{Department of Physics, Indiana University, Bloomington, Indiana 47405,
USA}

\author{G. Ortiz}
\affiliation{Department of Physics, Indiana University, Bloomington, Indiana 47405,
USA}
\affiliation{Department of Physics, University of Illinois, 1110 W Green Street, Urbana, Illinois 61801, USA}

\author{B. Seradjeh}
\affiliation{Department of Physics, Indiana University, Bloomington, Indiana 47405,
USA}

\author{E. Grosfeld}
\affiliation{Department of Physics, Ben-Gurion University of the Negev, Beer-Sheva
8410501, Israel}

\begin{abstract}
We study the quantum dynamics of Majorana and regular fermion bound states coupled to a quasi-one-dimensional metallic lead. The dynamics following the quench in the coupling to the lead exhibits a series of dynamical revivals as the bound state propagates in the lead and reflects from the boundaries. We show that the nature of revivals for a single Majorana bound state depends uniquely on the presence of a resonant level in the lead. When two spatially separated Majorana modes are coupled to the lead, the revivals depend only on the phase difference between their host superconductors. Remarkably, the quench in this case effectively performs a fermion-parity interferometry between Majorana bound states, revealing their unique non-Abelian braiding. Using both analytical and numerical techniques, we find the pattern of fermion parity transfers following the quench, study its evolution in the presence of disorder and interactions, and thus, ascertain  the fate of Majorana bound states in a Fermi sea.
\end{abstract}

\date{\today}

\maketitle

\emph{Introduction.}---%
One of the most intriguing features of a topological phase is the emergence of low-energy quasiparticles with fractional quantum numbers. Two examples of this fractionalization are solitons with fractional charge~\cite{SuSchHee79a,JacReb76a} and Majorana bound states (MBS) with non-Abelian exchange statistics~\cite{Ali12a,EllFra15a}. The MBSs are of particular interest due to their potential application in topological quantum computation~\cite{Kit03a}. While the existence of these fractional excitations has been proposed theoretically in many systems~\cite{ReaGre00a,Kit01a,WeeRosSer07a,HouChaMud07a,SerWeeFra08a,Fu2008,SatFuj09a,SatTakFuj09a,Lutchyn2010,OreRefOpp10a,Deng12,Ser12b,SerGro11a}, the experimental effort for their realization and detection is still the subject of vigorous current research. Recently, a number of groups have reported observing signatures of MBSs~\cite{Mourik2012a,DasRonMos12b,FinVanMoh13a,Nadj-Perge2014,Albrecht2016}. However, these experiments are spectroscopic in nature~\cite{LawLeeNg09a} and, thus, do not provide evidence of their non-Abelian statistics. In order to do so, one needs to perform a suitable interferometry~\cite{Akhmerov2009,clarke2010improved,GroSerVis11a,Akhmerov2010,GroSte11}, which is typically harder to do.

In this Rapid Communication, we propose a fermion-parity interferometry based on the quantum dynamics of bound states after a quench couples them to a metallic lead. The dynamics of the ground state after such a quench shows revivals~\cite{Quan2006,Cardy2014} at integer multiples of return time $\tau=2\ell/v_F$, where $\ell$ is the length and $v_F$ the Fermi velocity of the lead. Some aspects of such quench dynamics have been recently studied~\cite{hsu2009quantum,Vasseur,Rajak2014,Rajak2014a,Andraschko2014,hegde2015quench,Setiawan2015,He2016,Sacramento2016}. In our case, after the quench, the bound state leaks out to the lead on a time scale $1/\Gamma = 2\hbar E_F/\lambda^2$, where $E_F$ is the lead Fermi energy and $\lambda$ is the quenched coupling~\cite{Vasseur}. The resulting wave packet propagates in the lead at a velocity $\sim v_F$ and returns to the original position at time $\tau$. At this time it tunnels back to the original bound state, simultaneously balanced with the leakout, thus partially reviving the original state. For $\Gamma\tau\gg 1$, the revivals are the main aspect of the dynamics as the bound states shuttle back and forth along the lead. Thus they may be intuitively expected to provide a setting for such an interferometry on the bound states and reveal their exchange statistics.

Motivated by this observation, we consider a lead coupled to one or two such bound states and study the quantum dynamics after a quench in the tunneling amplitudes. Remarkably, we find a unique pattern of fermion parity transfers accompanying the revivals of MBSs, due to their nonlocal encoding of fermion parity. For regular fermion bound states (including Andreev bound states), fermion parity is encoded locally and no fermion parity transfers occur. We present analytical solutions for an effective low-energy theory of the quench dynamics of Majorana and regular fermion bound states. We also report numerical solutions to a full lattice model, which allow us to study the effects of potential disorder and local interactions in the lead. 

In all cases, we find unique dynamical signatures of MBSs. The quench dynamics of MBSs in the lead effectively performs a fermion-parity interferometry, revealing their non-Abelian braiding. In the presence of disorder and interactions in the lead, the amplitude of this pattern is eventually washed out after several return cycles. Nevertheless, the pattern of fermion parity transfers remains robust, thus providing a smoking gun for MBS detection.

\emph{Low-energy effective theory.}---%
Our interferometer is composed of a system (s) in a gapped phase hosting regular or Majorana bound states, and
a metallic lead (l) joined after a quench in a tunneling region (t), with the time-dependent Hamiltonian $H(t)=H_{\mathrm{s}}+H_{\mathrm{l}}+H_{\mathrm{t}}(t)$. The MBS $\gamma_{\nu a}$ ($\gamma_{\nu a}^2=\frac12$) at endpoint $a$ of superconductor $\nu$ contributes to the mode expansion of the electron operator the term $u_{\nu a}(x) \gamma_{\nu a}$, where the eigenfunction $u_{\nu a}(x)\propto e^{-x/\xi_\nu}e^{i(\phi_\nu+\eta_a) /2}$ with $x$ the position, $\xi_\nu$ the coherence length, $\phi_\nu$ the phase of the superconductor, and $\eta_a = 0\,(\pi)$ at the right (left) endpoint, $a=\I\,(\II)$.

For simplicity, we model the lead with spinless fermions, neglecting any spin dynamics. In the case of MBSs arising in spin-filtered nanowires~\cite{Kit01a,Lutchyn2010,OreRefOpp10a}, this is a good approximation as long as there are no magnetic impurities in the lead. In our low-energy theory, the lead degrees of freedom are the right- and left-moving modes, $\psi_R$ and $\psi_L$, with Hamiltonian  (in units of lattice spacing $\mathfrak{a}=\hbar=1$) $H_\text{l}=-iv_F\int_0^\ell(\psi^\dagger_R \partial_x \psi_R-\psi^\dagger_L \partial_x \psi_L)dx$. We now ``unwrap'' the lead coordinate $x\in[0,\ell]\mapsto[-\ell,\ell]$ by mapping $\psi_R(x)\mapsto\psi(x), \psi_L(x)\mapsto\psi(-x)$ to find
$
H_\text{l} = -i v_F \int_{-\ell}^\ell \psi^\dagger \partial_x \psi \, dx.
$
Equivalently, in the Majorana basis, $\psi=\frac1{\sqrt 2}(\gamma_1+i\gamma_2)$, we have
$
H_\text{l}= -\frac i2 v_F \int_{-\ell}^\ell (\gamma_1 \partial_x\gamma_1 + \gamma_2 \partial_x\gamma_2)\, dx.
$
Thus the lead is composed of two independent Majorana chains.
In this low-energy theory, the existence of a resonant zero-energy level is accounted for in the boundary condition $\psi(x+2\ell)=\zeta\psi(x)$, where $\zeta=+1\,(-1)$ when there is a (no) resonant level.

\begin{figure}
\includegraphics[width=3.35in]{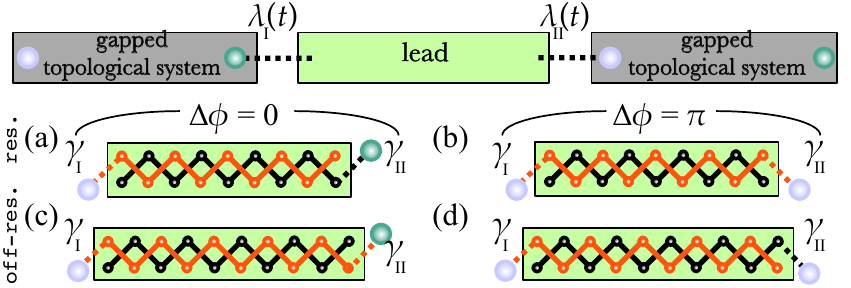}
\caption{ (Top) Schematic of the system; one or both tunnel couplings, $\lambda_{\I/\II}(t)$, are switched on at $t=0$. (Bottom) Coupling of Majorana bound states to the lead, shown in its Majorana basis as two independent Majorana chains. The resonant lead with odd number of sites [(a) and (b)] and the off-resonant lead with even number of sites [(c) and (d)] couple to Majorana bound states $\gamma_\I$ and $\gamma_\II$ in superconductors with phase difference $\Delta\phi= 0$ [(a) and (c)] and $\Delta\phi=\pi$ [(b) and (d)]. }\label{fig:leadMF}
\end{figure}

Since the system is gapped, at low energies we need only consider the contribution of the bound states to the tunneling Hamiltonian. Of course, all states in the superconductor couple to the lead; however, as we discuss in the conclusion, these states do not contribute to the physics of parity switching at return times. For regular fermions, $d_{\nu a}$, $H_\text{t}(t) = \sum_{\nu a}\lambda_{\nu a}(t) d_{\nu a}^\dagger\psi(x_a)+\text{H.c.}$, where $x_a$ is the tunneling site in the lead and $\lambda_{\nu a}$ the corresponding tunneling amplitude. 
For the MBS $\gamma_{\nu a}$, the tunneling term $H_\text{t}(t)$ in the lead Majorana basis is
\begin{equation}\label{eq:MFtunnel}
i\lambda_{\nu a} (t) \left[\sin \frac{\phi_\nu+\eta_a}2 \gamma_1(x_a) - \cos\frac{\phi_\nu+\eta_a}2 \gamma_2(x_a)\right] \gamma_{\nu a}.
\end{equation}
The quench is assumed to be sudden, $\lambda_{\nu a}(t)=\lambda_{\nu a}\theta(t)$.

The phase $\phi_\nu+\eta_a$ of one of the superconductors at a single contact point can always be gauged away by mapping $\psi\mapsto e^{-i(\phi_\nu+\eta_a)/2}\psi$. Thus, the dynamics only depends on the relative phase $\Delta\phi$ between the superconductors. 
Moreover, as illustrated in Fig.~\ref{fig:leadMF}, for a resonant lead coupled to two superconductors, the MBSs couple to the opposite (same) Majorana chains in the lead for $\Delta\phi=0\, (\pi)$. By contrast, for an off-resonant lead, the situation is reversed.
Thus, there are two cases to consider, in which one or two MBSs couple to a single lead Majorana chain. We shall now study these cases in detail.

\emph{MBS parity transfer.}---%
First, let us consider the lead coupled at $x_\I=0$ to a single MBS, $\gamma_\I\equiv\gamma$, with tunneling amplitude $\lambda$. The tunneling Hamiltonian is found by setting $\phi_\nu+\eta_a=0$ in Eq.~(\ref{eq:MFtunnel}) to be $i\lambda \gamma \gamma_2(0)$. Thus $\gamma_1$ is a free mode, and the other equations of motion are
\begin{align}
i\partial_t \gamma_2(x,t) &= -iv_F\partial_x\gamma_2(x,t) - i\lambda \gamma(t)\delta(x), \label{eq:EOM1}\\
i\partial_t \gamma(t) &= i\lambda\gamma_2(0,t). \label{eq:EOM2}
\end{align}
In order to account for revivals, we model the scattering off the bound state as a time-periodic perturbation. This is consistent with the linear continuum model since in this approximation all the lead modes propagate at the Fermi velocity, so they all scatter in regular intervals of return time. Thus, at $x=0^-$ and for $0<t<\tau$, we have a free field $\gamma_2(0^-,t)=\sum_\omega e^{-i\omega t}\gamma_{20}(\omega) \equiv \gamma_{20}(t)$, where $\gamma_{20}(\omega)$ are the modes of the unperturbed lead with energy $\omega$. The solution is given by 
\begin{equation}
\gamma(t)=f_{\lfloor{t/\tau}\rfloor}(t;\tau)\,\gamma(0) + F[\gamma_{20}],
\end{equation}
where $F$ is a functional of $\gamma_{20}$ only. Denoting $\Gamma=\lambda^2/2v_F$, the envelope functions 
$f_0(t;\tau) = e^{-\Gamma t}$ and, assuming $\Gamma\tau\gg1$,
$
f_1(t;\tau) = - 2 \zeta \Gamma (t-\tau)e^{-\Gamma(t-\tau)},
f_2(t;\tau) = -2\Gamma (t-2\tau)[1-\Gamma (t-2\tau)]e^{-\Gamma(t-2\tau)}
$.

The fermion parity of the host superconductor is $P(t)\equiv\langle 2i\gamma'\gamma(t) \rangle$, where $\gamma'$ is the spatially separated MBS partner of $\gamma$, which remains static. For $0<t<\tau$, $P(t) = e^{-\Gamma t} P(0)$, and is revived for $\tau<t<2\tau$ as
\begin{equation}
P(t)/P(0) = -2\zeta \Gamma(t-\tau)e^{-\Gamma(t-\tau)} , \quad \Gamma\tau\gg 1.
\end{equation}
The maximum revival value is $|P(\tau+1/\Gamma)/P(0)|= 2/e \approx 0.73$. Remarkably, the sign of the fermion-parity revival depends on $\zeta$: in a resonant lead it reverses. This pattern continues following each revival with the sign of the maximum parity switching at odd multiples of $\tau$.

For a regular fermion bound state, $d$, the equations of motion are found by replacing $\gamma_2\to\psi$, $\gamma\to id$ in~(\ref{eq:EOM1}) and~(\ref{eq:EOM2}). Similarly, the solution is $d(t) = f(t)\,d(0) - iF[\psi_0]$ with a free field $\psi_0(t)$. Thus the occupation of the bound state $N(t)\equiv\langle d^\dagger(t) d(t) \rangle = e^{-2\Gamma t} N(0)$ decays for $0<t<\tau$, and is revived for $\tau<t<2\tau$ as $N(t) \approx 4\Gamma^2 (t-\tau)^2e^{-2\Gamma(t-\tau)}$ with a maximum value $N(\tau+1/\Gamma)/N(0)=4/e^2\approx0.54$, irrespective of $\zeta$. The fermion parity of the host system, $P(t)=1-2N(t)$, is independent of $\zeta$.

Therefore the pattern of fermion parity transfers between the MBS and the lead upon revivals is a unique signature of MBSs. This is our first main result.

\emph{MBS fermion parity interferometry.}---%
It may appear too difficult to observe such a pattern of fermion parity transfers since tuning a lead level to be resonant requires a high degree of resolution. However, as we now show, the fermion parity transfers between two MBSs coupled to the lead is a robust signature of their non-Abelian exchange regardless of the nature of the lead. 

For simplicity, we will assume here the phase difference between the host superconductors $\Delta\phi=0$ or $\pi$ as in Fig.~\ref{fig:leadMF}. For $\Delta\phi=0$ in a resonant lead ($\zeta=+1$), and for $\Delta\phi=\pi$ in the off-resonant lead ($\zeta=-1$), the two MBSs $\gamma_\I$ and $\gamma_\II$ couple to different lead Majorana modes, $\gamma_1$ and $\gamma_2$. Thus our previous analysis shows that fermion parity is switched at return time $\tau$ only when $\zeta=+1$.

\begin{figure}[t] 
\includegraphics[scale=1]{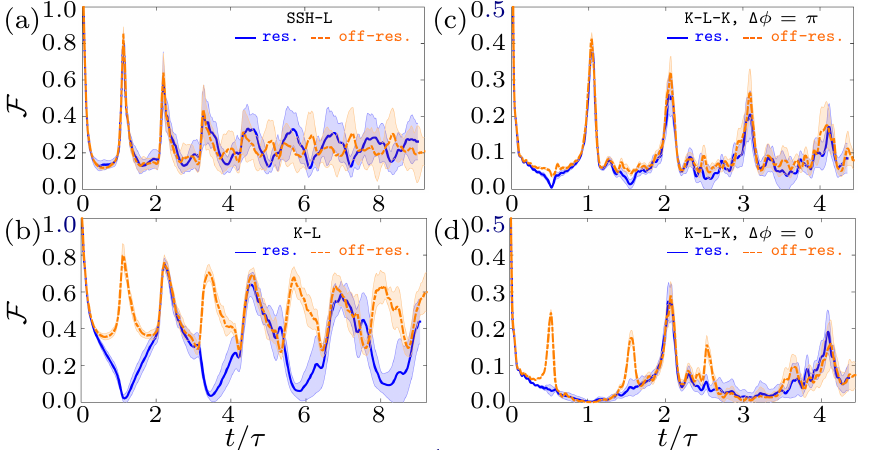}
\caption{Fidelity after the quench $\lambda=0.5w$ at $t=0$ couples a lead to
a single SSH chain with $m=0.8w$ (a), one Kitaev (b), and two Kitaev chains with gap $\Delta=0.3w$, $\Delta\phi=\pi$, (c) and $\Delta\phi=0$ (d). The disorder strength $W=0.2 w$ and the shaded areas show the standard deviation in the disorder average.}\label{fig:Fid1}
\end{figure}

The new cases are when both MBSs are coupled to the same lead Majorana mode, $\gamma_2$, i.e. $\Delta\phi=0$ in the off-resonant lead, and $\Delta\phi=\pi$ in the resonant lead. Then the equations of motion are
\begin{equation}
i\partial_t\gamma_2 = -iv_F\partial_x\gamma_2 - i\lambda_\I\gamma_\I\delta(x) - i\lambda_\II\gamma_\II\delta(x-\ell),
\end{equation}
and $i\partial_t\gamma_\I=i\lambda_\I\gamma_2(0)$, $i\partial_t\gamma_\II=i\lambda_\II\gamma_2(\ell)$. To proceed, we need to modify our previous calculation slightly to account for the scattering off the second MBS at odd multiples of $\tau/2$. This can be done straightforwardly, and for the simple case $\lambda_\I=\lambda_\II=\lambda$, the result is a decay for $0<t<\tau/2$, $\gamma_a(t)=f_0(t;\tau/2)\,\gamma_a(0) + F[\gamma_{20}]$, followed by a revival for $\tau/2<t<\tau$ (assuming $\Gamma\tau\gg 1$),
\begin{align}
\gamma_\I(t) &= \phantom{\zeta}f_1(t;\tau/2)\,\gamma_\II(0) + F[\gamma_{20}], \label{eq:gammaR} \\
\gamma_\II(t) &= \zeta f_1(t;\tau/2)\,\gamma_\I(0) + F[\gamma_{20}]. \label{eq:gammaL}
\end{align}
Thus the relative sign of exchange depends on $\zeta$. Projected to the subspace spanned by $\gamma_{R,L}$, the effective evolution operator for $\zeta=-1$ is $\sqrt{|f_1|}U_-$, where $U_-=e^{\frac\pi2\gamma_\II\gamma_\I}$ is the \emph{non-Abelian} Ising braid operator~\cite{ReaGre00a,Iva01a}. Hence the state of MBSs at $\tau/2$ is a superposition of two states in which \emph{both} MBS fermion parities are unchanged or switched. 
By contrast, 
for $\zeta=+1$, the projected evolution operator is $\sqrt{|f_1|}U_+$, where $U_+=\gamma_\II+\gamma_\I$ satisfies $U_+^2=1$ as in an \emph{Abelian} braiding. In this case, the state of MBSs at $\tau/2$ is a superposition of two states in which one or the other MBS switches its fermion parity. 
Proceeding to $\tau<t<3\tau/2$, we find $\gamma_a(t)=\zeta f_2(t;\tau/2)\,\gamma_a(0)+F$: upon revival, fermion parities switch only for $\zeta=-1$.

We conclude that at odd multiples of return time, the fermion parities of $\gamma_\I$ and $\gamma_\II$ are (not) switched, regardless of the details of the lead, when $\Delta\phi=0$ ($\pi$). This is our second main result.

\emph{Numerics and effects of disorder and interactions.}---%
In order to confirm and extend our results beyond the clean, noninteracting low-energy limit, we now study the quench dynamics of the many-body system in a lattice model numerically. 

We model the system as a one-dimensional chain with Hamiltonian
$
H_{\mathrm{s}} = \sum_{s}\left(w_s d_{s}^{\dagger}d_{s+1}+\Delta d_{s}d_{s+1}+\mbox{H.c.}\right),
$
where $d^\dagger_s$ is the system fermionic creation operator at site $s$, $w_s= w+(-1)^s m$ is the hopping amplitude with $m$ the bond modulation, and $\Delta=|\Delta| e^{i\phi}$ is the superconducting pairing. This Hamiltonian includes the Su-Schrieffer-Heeger (SSH) model~\cite{SuSchHee79a} with $m\neq0,\Delta=0$, which supports regular fermion bound states (solitons), and the Kitaev model~\cite{Kit01a} with $m=0,\Delta\neq 0$, which supports MBSs, at the chain's endpoints. 
We take
\begin{align}
H_{\mathrm{l}}
	&= w\sum_{r=1}^{N-1}\left(c_{r}^{\dagger}c_{r+1}+\mbox{H.c.}\right)+\sum_{r=1}^N V_r\left(n_r-\frac{1}{2}\right) \nonumber\\
	& + U\sum_{r=1}^{N-1}\left(n_r-\frac{1}{2}\right)\left(n_{r+1}-\frac{1}{2}\right),
	\label{eq:lead}
\end{align}
for the lead, where $c^\dagger_r$ is the fermionic creation operator at site $r$, $n_r=c_r^\dag c_r$ is the number operator, $U$ the interaction strength, and $V_r$ the potential disorder with a uniform distribution over $[-W/2,W/2]$ and disorder strength $W$. The lead is (off-)resonant for (even) odd $N$.
The tunneling Hamiltonian is
$
H_{\mathrm{t}}(t)=\sum_{rs}{\lambda}_{rs}(t) d_{s}^{\dagger} c_{r}+\mbox{h.c.},
$
with $\lambda_{rs}(t)$ the quenched tunneling amplitude between the system
site $s$ and the lead site $r$.

\begin{figure*}[t!!]
\includegraphics[scale=1.02]{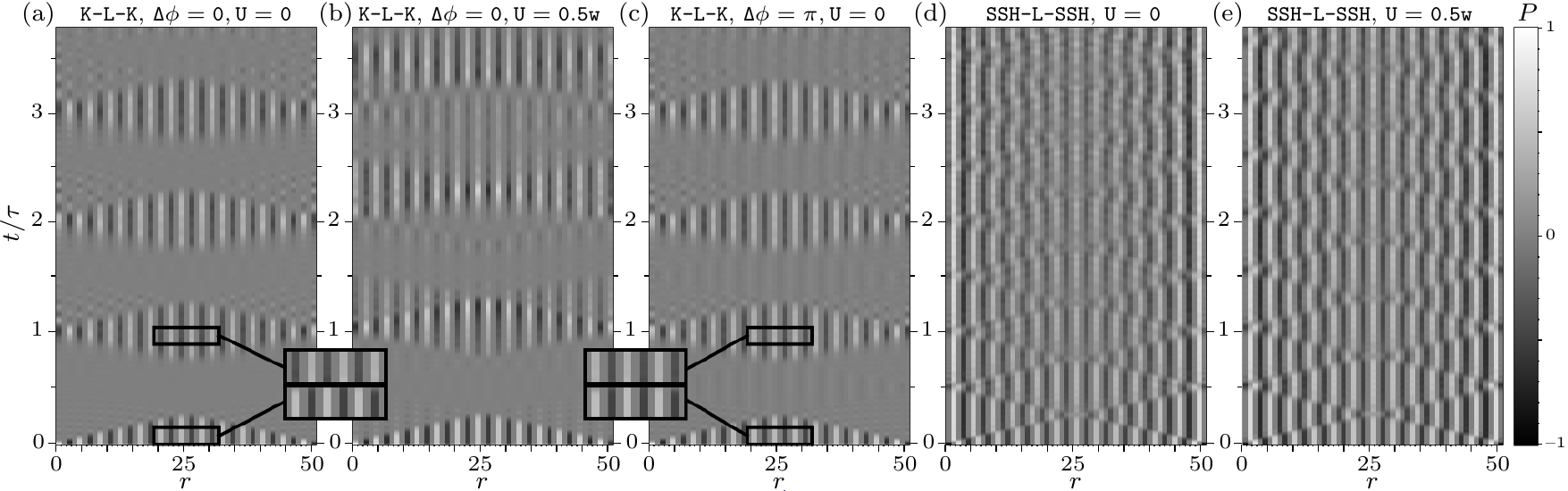}
\caption{The parity $P(r,t)$ to the left of a cut between sites $r$ and $r+1$ in the lead after a quench $\lambda=w$ at $t=0$ couples the lead to two Kitaev chains with gap $\Delta=0.5w$ and $\Delta\phi=0$ [(a) and (b)], $\Delta\phi=\pi$ (c), and to two SSH chains with $m=0.8w$ [(d) and (e)]. The lead interaction strength $U=0.5w$ in (b) and (e). The insets show magnified regions for comparison.}\label{fig:Par2}
\end{figure*}

As a first measure of the quench dynamics, we compute the dynamical fidelity, or the Loschmidt echo~\cite{Jalabert2001,Gorin2006,Gu2010},
$
	\mathcal{F}(t)=|\langle \Psi(0)|\Psi(t)\rangle|^2,
$
where $|\Psi(t)\rangle$ is the many-body ground state of the system. For this calculation, we set $U=0$ and diagonalize the Hamiltonian exactly. The ground state overlaps can then be calculated using the Onishi formula~\cite{Onishi1966,Balian1969}. 

A sampling of our results is shown in Fig.~\ref{fig:Fid1} for a disorder strength $W=0.2w$. For a single soliton, Fig~\ref{fig:Fid1}(a), the fidelity is revived at integer multiples of $\tau$. However, for a single MBS and a resonant lead, Fig.~\ref{fig:Fid1}(b), the fidelity exhibits a dip at odd multiples of $\tau$. This is in agreement with the pattern of fermion parity transfers, since fermion parities are switched. At disorder strength larger than the level spacing, the distinction starts to disappear; however, the MBS even-odd effect can still be seen in the first few revivals since the signal is averaged over a minimum and a maximum at odd return times.

For two MBSs, Figs.~\ref{fig:Fid1}(c) and~\ref{fig:Fid1}(d), our numerics again confirm the even-odd effect at return time: in contrast to $\Delta\phi=\pi$, the parities are switched at odd multiples of $\tau$ for $\Delta\phi=0$ and, instead of a maximum, the fidelity shows a dip. The structure around $\tau/2$ depends on the resonant nature of the lead: or MBSs with $\Delta\phi=\pi$ coupled to a resonant lead, the total fermion parity is switched at $\tau/2$ and, thus,  we expect fidelity to show a dip; however, for MBSs with $\Delta\phi=0$ coupled to an off-resonant lead, the state at $\tau/2$ has an amplitude $\sqrt{|f_1|/2}$ to be in the original state, yielding a maximum fidelity $1/e$, consistent with our numerics. We note that MBSs are more robust against potential disorder than the SSH solitons. We have confirmed that the revival pattern of MBSs in fidelity remains robust over several return cycles for relatively high disorder strengths, $W\lesssim 0.5 w$.


Indeed, the pattern of fermion parity transfers can be directly observed in our numerics. As a second measure, we calculate the fermion parity, $P(r,t)=\langle \Psi(t)| (-1)^{N_r} |\Psi(t) \rangle$, where $N_r$ is the total number operator to the left of a cut between sites $r$ and $r+1$ in the lead. Since the total fermion parity is conserved, $P(r_a,t)$ for the case with two tunneling contacts at endpoints $r_a$ directly measures the fermion parity of the systems hosting the bound states at endpoint $a$. For this calculation we use the
time-dependent density-matrix renormalization group method~\cite{white1992density,white2004real}, which allows us to study the effects of disorder and interactions. For a single bound state, we have checked that the fermion parity is switched at odd multiples of $\tau$ only for the single MBS coupled to the resonant lead.

Figure~\ref{fig:Par2} summarizes our numerical results for fermion parity transfers for two bound states. In agreement with our analytical solutions, the fermion parity of the MBSs is switched at odd multiples of $\tau$ when $\Delta\phi=0$ and not so when $\Delta\phi=\pi$ regardless of the nature of the lead. By contrast, the fermion parity of the SSH solitons does not switch. For $U=0.5w$, the pattern of fermion parity transfers for the MBSs decoheres over a few return cycles. The SSH solitons, on the other hand, are much less affected. This is consistent with the idea that the fermion parity transfers manifest subtle interference paths of MBSs, which are more prone to decoherence by local interactions. We note, however, that the same switching pattern is observed as in the noninteracting case, albeit in an increasingly incoherent fashion.

\emph{Concluding remarks.}---%
Our fermion parity interferometry can distinguish MBSs from other fermionic bound states, including Andreev bound states. To see this, note that fermion parity transfers occur due to the nonlocal encoding of fermion parities in spatially separated MBSs. For an Andreev bound state $\gamma_E$ at energy $E$, the fermion parity $\sim\langle\gamma_E^\dagger\gamma_E\rangle$  is local. Thus, like regular fermions, Andreev bound states would not show fermion parity transfers in our scheme. We have indeed confirmed this numerically.


We also considered a lead with multiple channels~\cite{suppl} and confirmed that as long as the notion of a return time is meaningful, the pattern of fermion parity transfers continues to hold.

Spatially separated MBSs have a tunneling time $1/\epsilon$ with energy splitting $\epsilon\sim e^{-\ell/\xi}$. Thus, our fermion parity interferometry would work for $1/\Gamma \ll \tau \ll 1/\epsilon$. The upper limit is not challenging since  $\ell/\xi$ can be large. Restoring units of $\ell$ in micron, $v_F$ in eV\AA, $\lambda$ in meV, and $\mathfrak{a}$ in \AA, the time scales are $1/\Gamma \sim (v_F/\lambda^2 \mathfrak{a})\times10^{-9}$s and $\tau \sim (\ell/v_F)\times 10^{-11}$s; thus, $\lambda \gtrsim \lambda_*= 10 v_F/\sqrt{\ell\mathfrak{a}}$. These values can vary significantly depending on the realization scheme. For typical solid-state parameters, $\lambda_*\sim10$meV, but it can be lowered for smaller $v_F$ and larger $\mathfrak{a}$. For example, in nanopatterned metallic surfaces~\cite{GomMarKo12a}, $\lambda_*<1$meV can be easily achieved. In cold-atom realizations~\cite{micheli2004single,Kraus2012,BakGilPen09a}, with $v_F\sim10^{-2}$cm/s, $\mathfrak{a}\sim100$nm, and $\lambda\sim1$kHz, we have $1/\Gamma\sim10^{-4}$s $\ll\tau\sim10^{-2}$s. We further discuss experimental feasibility in Ref.~\cite{suppl}.


We have investigated the quench dynamics of a topological system coupled to a Fermi sea. We have found that the lead can serve as an interference medium revealing the non-Abelian exchange statistics of MBSs through a unique pattern of fermion parity transfers. Remarkably, this pattern remains the same in the presence of moderate interactions and disorder in the lead. We note that unlike effective braiding of MBSs~\cite{HecAkhHas12a}, the exchange in our setup proceeds via the real-space paths of the lead channels. Our findings can lead to viable interferometers for the smoking-gun detection of Majorana bound states.

\begin{acknowledgments}
\emph{Acknowledgments}---%
We thank E. Fradkin for useful discussions. E.G. acknowledges support from the Israel Science Foundation under Grants
No. 401/12 and  No. 1626/16, and the European Union's Seventh Framework Programme (FP7/2007-2013)
under Grant No. 303742. E.G. and B.S. acknowledge support from the Binational
Science Foundation through Grant No. 2014345. D.D thanks the support of H.B.C. fellowship (Israel). This work is supported in part by the NSF CAREER Grant  No. DMR-1350663 as well as the College of Arts and Sciences at Indiana University (M.T.A. and B.S.). We also thank the hospitality of Aspen Center for Physics, supported by National Science Foundation Grant  No. PHY-1066293, where parts of this work were performed.
\end{acknowledgments}

\end{bibunit}

\begin{bibunit}

\section{Supplemental Material}

In this Supplemental Material we provide additional details regarding the quench dynamics for Majorana fermions suddenly coupled to multiple Fermi liquid channels. We demonstrate analytically and numerically that the revival pattern is robust in the multi-channel case over several cycles of revivals. We also present a discussion of the feasibility of detecting our findings experimentally.

\subsection{Multiple channels}
We consider $M$ independent lead channels and couple them to either a single Majorana bound state one end, or to two at both ends. The tunneling matrix element in each channel is $w_\alpha$ ($\alpha=1,\ldots,M$), and the tunneling matrix element to the Kitaev chain is $\lambda_\alpha$. Then, for a coupling to Majorana bound state $\gamma_a$, $H_\text{t} = \sum_\alpha \lambda_{\alpha a} \gamma_a c_{\alpha a}+\text{h.c}$. The lead Hamiltonian $H_\text{l}=\sum_\alpha w_\alpha c^\dagger_{\alpha r}c_{\alpha r+1}+\text{h.c}$.

\subsubsection{Degenerate channels}
The tunneling Hamiltonian can always be written as $H_\text{t}=|\lambda_a| \gamma_a \tilde c_{1a}$ via a rotation in the channel basis, $\tilde c_{1a} = \sum_\alpha \lambda_{\alpha a} c_{\alpha a}/|\lambda_a|$, where $|\lambda_a| = \sqrt{\sum_\alpha \lambda_{\alpha a}^2}$. This rotation can be uniformly extended to all sites in the lead by a unitary transformation, $U$, in the channel basis, $\tilde c_{\alpha r} = \sum_{\beta} U_{\alpha\beta} c_{\beta r}$, such that $U_{1\beta} = \lambda_{\beta a}/|\lambda_a|$. For degenerate channels $w_\alpha = w$, and the lead Hamiltonian in the new basis is also $H_\text{l} = \sum_\alpha w_\alpha \tilde c^\dagger_{\alpha r} \tilde c_{\alpha r+1}+\text{h.c}$. Thus, for degenerate channels, the problem is reduced to a Majorana bound state coupled to a single (rotated) channel in the lead. Therefore, our results remain exact.

\subsubsection{Non-degenerate channels}
For non-degenerate channels, we return to the effective low-energy theory. Then, the lead Hamiltonian in the Majorana basis is given by
\begin{eqnarray}
H_{\mathrm{l}}=-\frac{i}{2}\int_{-\ell}^{\ell} \sum_{\alpha} v_{F \alpha}(\gamma_{1\alpha}\partial_x\gamma_{1\alpha}+\gamma_{2\alpha}\partial_x\gamma_{2\alpha})dx,
\end{eqnarray}
and, the tunneling Hamiltonian is, for $t>0$,
\begin{eqnarray}
H_{\mathrm{t}}=i\sum_{\alpha}\lambda_\alpha\gamma\gamma_{2\alpha}(0).
\end{eqnarray}
We shall drop the endpoint index $a$ here for brevity.

The solution of the equations of motion results in the following relation
\begin{eqnarray}
(\partial_t+\Gamma)\gamma(t)=\sum_\alpha \lambda_\alpha\gamma_{2\alpha}(0^-,t) \equiv |\lambda| \gamma_{20}(t),
\end{eqnarray}
where, now $\Gamma=\sum_\alpha \lambda_\alpha^2/2v_{F\alpha}$ and the free field $\gamma_{20}(t) = \sum_\alpha \lambda_{\alpha} \gamma_{2\alpha}(0^-,t)/|\lambda|$. This equation has the same form as the one leading to the solution in Eq.~(4) in the main text. Thus, as long as the channels have return times, $\tau_\alpha = 2\ell/v_{F\alpha}$, that are close enough to a common return time, $\tau$, we obtain the same revival pattern as in the case of a single channel. More precisely, define $\tau_< = \min_\alpha(\tau_\alpha)$, and $ \tau_> = \max_\alpha(\tau_\alpha)$; then, if after $n$ revivals, $n \tau_< \gg  (n-1)\tau_>+\frac{1}{\Gamma}$, the solution for $\gamma$ has the same form in the first $n-1$ revivals. Thus, as long as the notion of a similar return time $\tau$ is well defined, i.e. the return times of different channels do not overlap, the same dynamics of revivals will result from the quench coupling to multiple channels. 

\subsubsection{Numerics with independent multiple channels}
\begin{figure*}
	\includegraphics[width=4.5in]{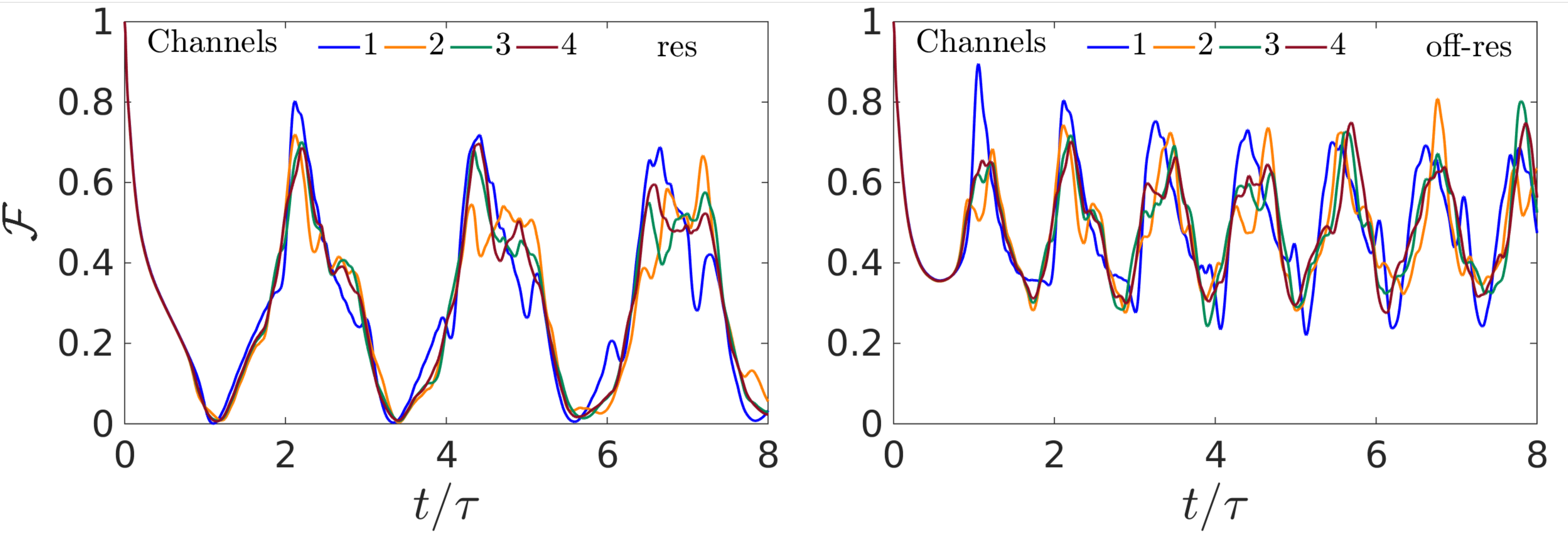}
	\caption{Fidelity for $M$ lead channels coupled to a single Kitaev chain on their end, as a function of time. Each channel contains $51$ (resonant, left panel) or $50$ (off-resonance, right panel) sites, and each Kitaev chain is composed of $50$ sites. Here $\lambda_\alpha=0.5/\sqrt{M}$, the gap $\Delta=0.3$ in the Kitaev chain and $w_\alpha$ are equally spaced within $[w-\delta w/2,w+\delta w/2]$ with $w=1$ and $\delta w=0.3$.}\label{fig:multi-channel-1}
\end{figure*}

\begin{figure*}
	\includegraphics[width=4.5in]{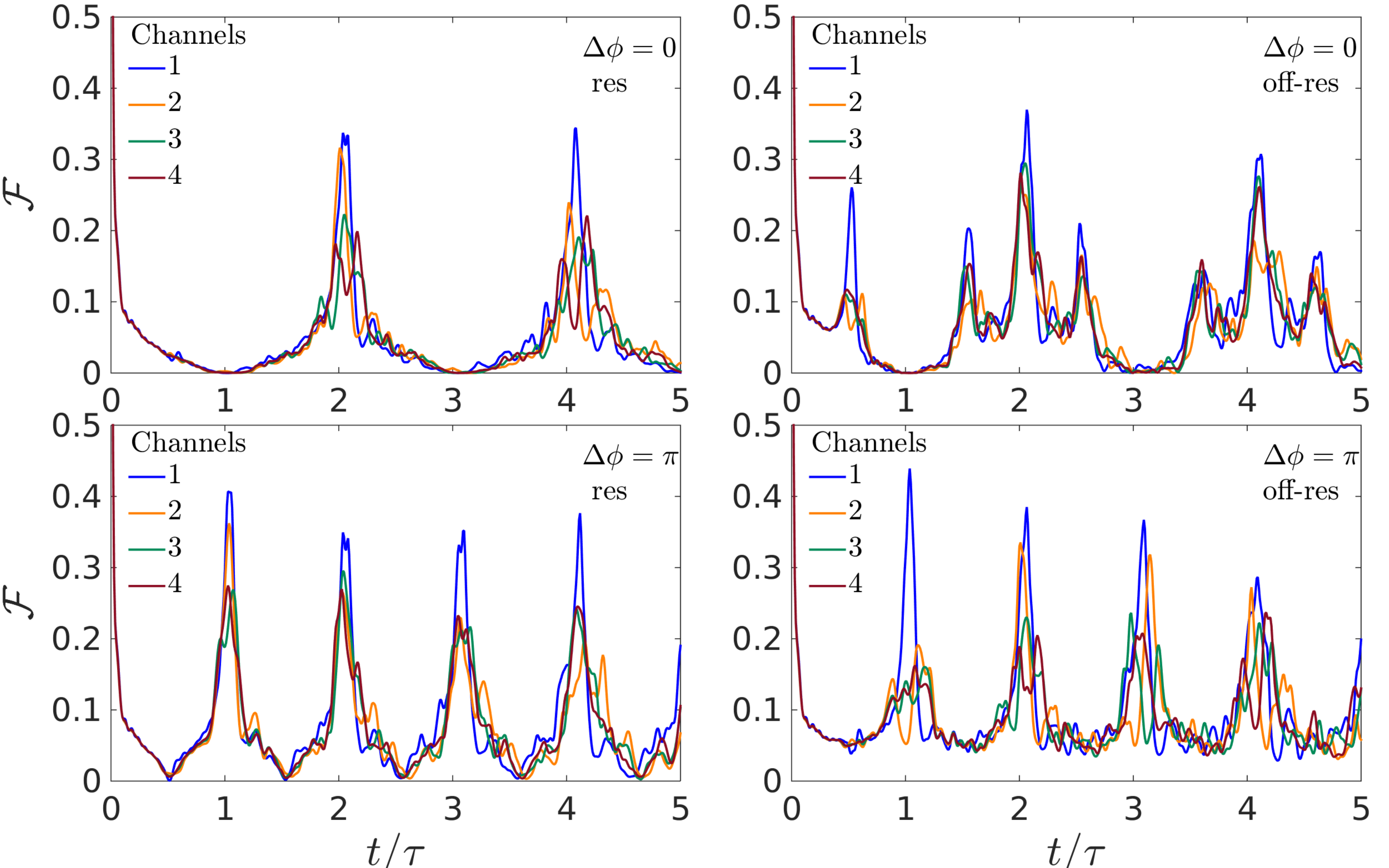}
	\caption{Fidelity for $M$ fermi liquid channels coupled to two Kitaev wires on their two ends, as function of time. Each channel contains $51$ (resonant, left panels) or $50$ (off-resonance, right panels) sites, and each Kitaev chain is composed of $50$ sites. Here $\lambda_\alpha=1/\sqrt{M}$, the gap $\Delta_\text{I}=0.3$ and $\Delta_\text{II}=0.3 e^{i\Delta \phi}$, and $w_\alpha$ are equally spaced within $[w-\delta w/2,w+\delta w/2]$ with $w=1$ and $\delta w=0.3$.}\label{fig:multi-channel-2}
\end{figure*}

We have performed numerical calculations to support the argument above, presented in Fig.~\ref{fig:multi-channel-1} for a single Kitaev chain and in Fig.~\ref{fig:multi-channel-2} for two Kitaev chains. We find that the suppression of revival peaks with the number of channels is weak while the appearance of a near zero in the fidelity remains robust. 

\begin{figure*}
	\includegraphics[width=4.5in]{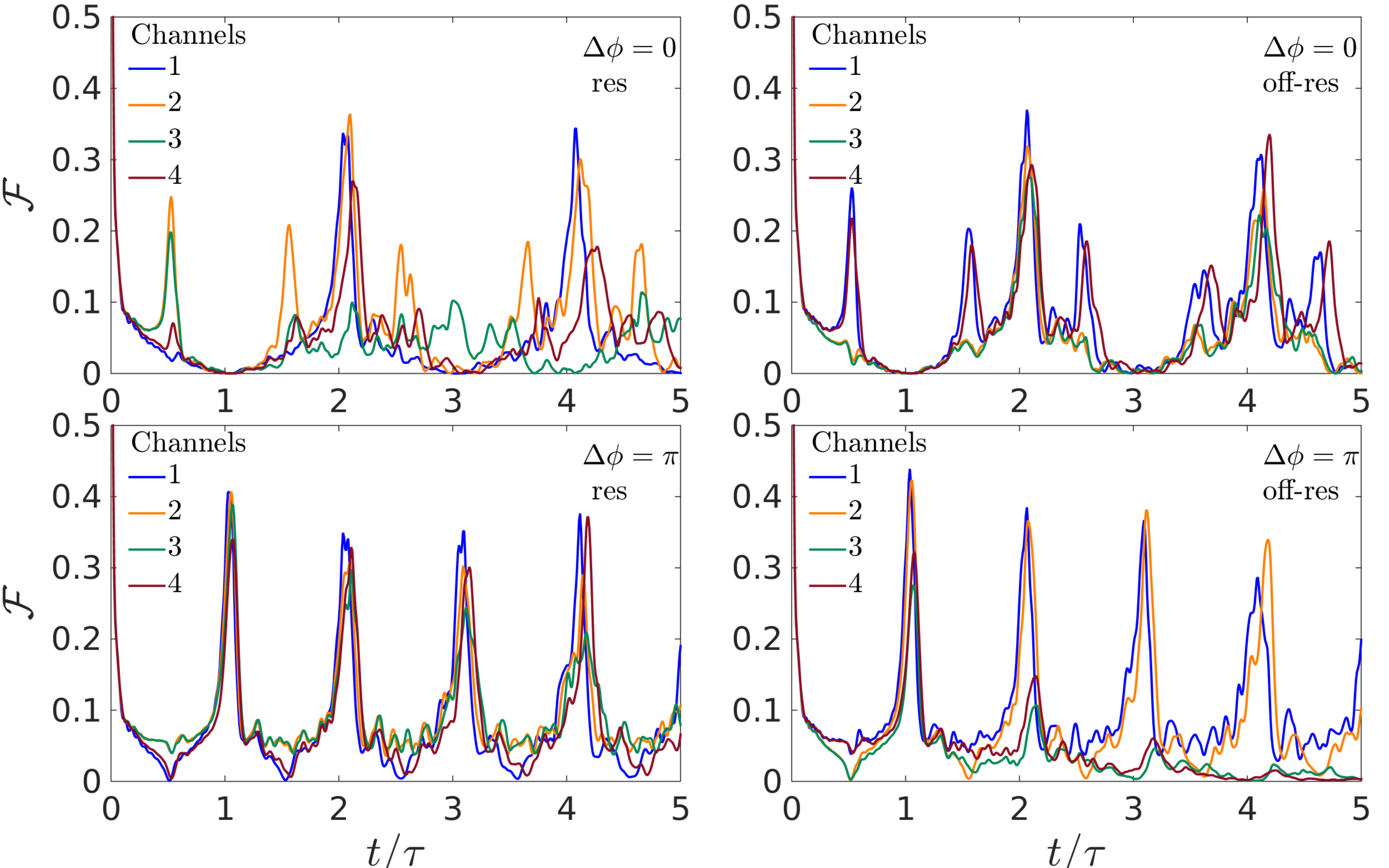}
	\caption{Fidelity for a lead, modeled as a $N\times M$ slab, coupled to two Kitaev wires on its two ends, as function of time. Here $\lambda_\alpha=1/\sqrt{M}$, the tunneling $w_\leftrightarrow=1.0$ between sites $r=1\cdots N$, and $w_\updownarrow=0.3$ between channels $\alpha=1\cdots M$, and the gaps $\Delta_\text{I}=0.3$ and $\Delta_\text{II}=0.3 e^{i\Delta\phi}$ in the Kitaev chain. The lead has $N=51$ (resonant, left panels) or $N=50$ (off-resonance, right panels).}\label{fig:multi-channel-3}
\end{figure*}

\subsubsection{Numerics on a slab}
We also performed a numerical study where the lead is modeled as a slab of $N\times M$ sites, with tunneling $w_{\leftrightarrow}$ on horizontal bonds and tunneling $w_{\updownarrow}$ on vertical bonds. The lead is connected to Kitaev chains on its left and right ends. The results for the fidelity are presented in Fig.~\ref{fig:multi-channel-3}. Also in this case, the dip at $t=\tau$ for $\phi=0$ (and the peak for $\phi=\pi$) shows considerable robustness in the presence of multiple non-degenerate channels. Again, the pattern of maxima and minima is the same as in the single-channel case for the first few cycles .

\subsection{Experimental feasibility}
In the main paper, we focused on revealing the basic physics in our proposed setup, and estimated time scales that suggest the experimental possibility of observing our findings in solid-state realizations ($\sim10^{-10}$ s) and a more accessible parameter regime in cold-atom systems ($\sim 10^{-2}$ s). We must emphasize that our proposal is not limited to measuring the Loschmidt echo, as, for example is the case in Ref.~\onlinecite{Vasseur}. Measuring a power-law decay over a smaller time scale~\cite{Vasseur} $1/\Gamma\ll\tau$ presents a tougher challenge than observing fermion parity exchanges over longer return time cycles, $\tau$. In this context, it is important to note also that, in the presence of multiple lead channels, the power-laws of Ref.~\onlinecite{Vasseur} are corrected and not universal. However, the pattern of revivals and fermion-parity exchanges in our setup remains the same. This additional robustness should make the observation of the physics we discuss more feasible.

In the following, we briefly discuss what could be a readily available way to observe our findings. Instead of requiring sensitivity to two different power-law decays~\cite{Vasseur}, our setup requires only to differentiate between (i) a zero in the fidelity (the near orthogonality of two wavefunctions), and (ii) a revival of the fidelity (existence of a finite overlap) due to fermion parity switchings. This can be done using fermion-parity sensing. Since fermion parity is a global property, this measurement need not be done locally or at the point of contact. Take a Kitaev-Lead-Kitaev setup and perform the following protocol:

\begin{enumerate}[(1)]
	\item Measure the fermionic parity of each of the two Kitaev wires prior to the quench.
	\item Turn on the couplings and wait for a time $\tau$. (If required by the time scales for achieving the next step, turn off the couplings between the subsystems now.)
	\item Measure the fermionic parity present at each of the two Kitaev wires following the quench.
\end{enumerate}

By comparing the results of steps (1) and (3) one can deduce whether the state after time $\tau$ is orthogonal or overlapping with the initial state. The experiment should then be repeated many times to ascertain a parity switch for each realization. Achieving steps (1) and (3) is within reach as measurement of single atoms in an optical lattice was already demonstrated experimentally~\cite{BakGilPen09a} and measurement of fermion parity is a byproduct. In addition, in comparison with other proposed setups for measuring non-Abelian statistics, our setup requires no complicated gating and does not rely on the adiabaticity of the exchange process.

\end{bibunit}	


\begin{thebibliography}{10}

\bibitem{SuSchHee79a}
W.~P. Su, J.~R. Schrieffer, and A.~J. Heeger, 
  \prl\ \textbf{42}, 1698 (1979).

\bibitem{JacReb76a}
R.~Jackiw and C.~Rebbi,
  \prd\ \textbf{13}, 3398 (1976).
  
\bibitem{Ali12a}
J.~Alicea, 
  {Rep. Prog. Phys.} \textbf{75}, 76501 (2012).

\bibitem{EllFra15a}
S.~R. Elliott and M.~Franz,
  \rmp\ \textbf{87}, 137 (2015).

\bibitem{Kit03a}
A.~Yu. Kitaev, 
  {Ann. Phys.} \textbf{303}, 2 (2003).

\bibitem{ReaGre00a} 
N.~Read and D.~Green,  
 \prb\ \textbf{61}, 10267 (2000). 

\bibitem{Kit01a}
A.~Yu. Kitaev,
  {Phys.-Usp.} \textbf{44}, 131 (2001).

\bibitem{WeeRosSer07a}
C.~Weeks, G.~Rosenberg, B.~Seradjeh, and M.~Franz, 
{Nature Phys.} \textbf{3}, 796 (2007).

\bibitem{HouChaMud07a}
C.-Y. Hou, C.~Chamon, and C.~Mudry,
  \prl\ \textbf{98}, 186809 (2007),

\bibitem{SerWeeFra08a}
B.~Seradjeh, C.~Weeks, and M.~Franz, 
  \prb\ \textbf{77}, 033104 (2008).

\bibitem{Fu2008}
L.~Fu and C.~L. Kane, 
  \prl\ \textbf{100}, 096407 (2008).

\bibitem{SatFuj09a} 
M.~Sato and S.~Fujimoto, 
 \prb\ \textbf{79}, 094504 (2009). 
 
\bibitem{SatTakFuj09a} 
M.~Sato, Y.~Takahashi, and S.~Fujimoto, 
 \prl\ \textbf{103}, 020401 (2009). 
 
\bibitem{Lutchyn2010}
R.~M. Lutchyn, J.~D. Sau, and S.~{Das Sarma}, 
  \prl\ \textbf{105}, 077001(2010).

\bibitem{OreRefOpp10a}
Y.~Oreg, G.~Refael, and F.~von Oppen,
  \prl\ \textbf{105}, 177002 (2010).

\bibitem{SerGro11a}
B.~Seradjeh and E.~Grosfeld, 
  \prb\ \textbf{83}, 174521 (2011).

\bibitem{Deng12}
S.~Deng, L.~Viola, and G.~Ortiz, 
  \prl\ \textbf{108}, 036803 (2012). 

\bibitem{Ser12b}
B.~Seradjeh,
  \prb\ \textbf{86}, 121101(R) (2012).

\bibitem{Mourik2012a}
V.~Mourik, K.~Zuo, S.~M. Frolov, S.~R. Plissard, E.~P. a.~M. Bakkers, and L.~P.
  Kouwenhoven, 
  {Science} \textbf{336}, 1003 (2012).

\bibitem{DasRonMos12b}
A.~Das, Y.~Ronen, Y.~Most, Y.~Oreg, M.~Heiblum, and H.~Shtrikman, 
  {Nature Phys.} \textbf{8}, 887 (2012).

\bibitem{FinVanMoh13a}
A.~D.~K. Finck, D.~J. {Van Harlingen}, P.~K. Mohseni, K.~Jung, and X.~Li,
  \prl\ \textbf{110}, 126406 (2013).

\bibitem{Nadj-Perge2014}
S.~Nadj-Perge, I.~K. Drozdov, J.~Li, H.~Chen, S.~Jeon, J.~ Seo, A.~H. MacDonald, B.~A. Bernevig, and A.~Yazdani,
  {Science} \textbf{346}, 602 (2014).
  
\bibitem{Albrecht2016} 
S.~M. Albrecht, A.~P. Higginbotham, M.~Madsen, F.~Kuemmeth, T.~S. Jespersen, J.~Nyg{\aa}rd, P.~Krogstrup, and C.~M. Marcus, 
  Nature \textbf{531}, 206 (2016). 
  
\bibitem{LawLeeNg09a} 
K.~T. Law, P.~A. Lee, and T.~K. Ng, 
 \prl\ \textbf{103}, 237001 (2009). ?
\bibitem{Akhmerov2009}
A.~R. Akhmerov, J.~Nilsson, and C.~W.~J. Beenakker, 
  \prl\ \textbf{102}, 216404 (2009).

\bibitem{clarke2010improved} 
D.~J. Clarke and K.~Shtengel, 
  \prb\ \textbf{82}, 180519(R) (2010). 
  
\bibitem{GroSerVis11a}
E.~Grosfeld, B.~Seradjeh, and S.~Vishveshwara,
  \prb\ \textbf{83}, 104513 (2011).
  
\bibitem{Akhmerov2010}
F.~Hassler, a.~R. Akhmerov, C.~Y. Hou, and C.~W.~J. Beenakker, 
  New J. Phys. \textbf{12}, 1 (2010).
  
\bibitem{GroSte11}
E.~Grosfeld and A.~Stern, 
  Proc. Natl. Acad. Sci. \textbf{108}, 11810 (2011).

\bibitem{Quan2006}
H.~T. Quan, Z.~Song, X.~F. Liu, P.~Zanardi, and C.~P. Sun,
  \prl\ \textbf{96}, 140604 (2006).

\bibitem{Cardy2014}
J.~Cardy, 
  \prl\ \textbf{112}, 220401 (2014).

\bibitem{hsu2009quantum} 
B.~Hsu, E.~Grosfeld, and E.~Fradkin, 
  \prb\ \textbf{80}, 235412 (2009). 

\bibitem{Vasseur}
R.~Vasseur, J.~P. Dahlhaus, and J.~E. Moore, 
  Phys. Rev. X \textbf{4}, 041007 (2014).

\bibitem{Rajak2014}
A.~Rajak and A.~Dutta, 
  \pre\ \textbf{89}, 042125 (2014).

\bibitem{Rajak2014a}
A.~Rajak, T.~Nag, and A.~Dutta, 
  \pre\ \textbf{90}, 042107 (2014).

\bibitem{Andraschko2014}
F.~Andraschko and J.~Sirker, 
  \prb\ \textbf{89}, 125120 (2014).

\bibitem{hegde2015quench} 
S.~Hegde, V.~Shivamoggi, S.~Vishveshwara, and D.~Sen, 
  N. J. Phys. \textbf{17}, 053036 (2015). 

\bibitem{Setiawan2015}
F.~Setiawan, K.~Sengupta, I.~B. Spielman, and J.~D. Sau, 
  \prl\ \textbf{115}, 190401 (2015).

\bibitem{He2016}
Y.~He and C.-C. Chien, 
  \prb\ \textbf{94}, 024308 (2016).

\bibitem{Sacramento2016}
P.~D. Sacramento, 
  \pre\ \textbf{93}, 062117 (2016).


\bibitem{Iva01a}
D.~A. Ivanov, 
  \prl\ \textbf{86}, 268 (2001).

\bibitem{Jalabert2001} 
R.~A. Jalabert and H.~M. Pastawski, 
  \prl\ \textbf{86}, 2490 (2001).

\bibitem{Gorin2006}
T.~Gorin, T.~Prosen, T.~H. Seligman, and M.~{\v{Z}}nidari{\v{c}}, 
  {Phys. Rep.} \textbf{435}, 33 (2006).
  
\bibitem{Gu2010}
S.-J. Gu, 
  {Int. J. Mod. Phys. B} \textbf{24}, 4371 (2010).

\bibitem{Onishi1966}
N.~Onishi and S.~Yoshida, 
  {Nucl. Phys.} \textbf{80}, 367 (1966).

\bibitem{Balian1969}
R.~Balian and E.~Brezin, 
  {Il Nuovo Cim. B} \textbf{64}, 37 (1969).

\bibitem{white1992density}
S.~R. White, 
  \prl\ \textbf{69}, 2863 (1992).

\bibitem{white2004real}
S.~R. White and A.~E. Feiguin, 
  \prl\ \textbf{93}, 076401 (2004).
  
\bibitem{suppl}  See Supplemental Information.
  
\bibitem{GomMarKo12a} 
K.~K. Gomes, W.~Mar, W.~Ko, F.~Guinea, and H.~C. Manoharan, 
  Nature \textbf{483}, 306 (2012). 

\bibitem{micheli2004single} 
A.~Micheli, A.~J. Daley, D.~Jaksch, and P.~Zoller,
  \prl\ \textbf{93}, 140408 (2004).  

\bibitem{Kraus2012}
C.~V. Kraus, S.~Diehl, P.~Zoller, and M.~A. Baranov, 
  {N. J. Phys.} \textbf{14}, 113036 (2012).
  
\bibitem{BakGilPen09a} W.~S. Bakr, J.~I. Gillen, A.~Peng, S.~Folling, and M.~Greiner, 
  {Nature} \textbf{462}, 74 (2009).
  
\bibitem{HecAkhHas12a} 
B.~van Heck, A.~R. Akhmerov, F.~Hassler, M.~Burrello, and C.~W.~J. Beenakker, 
  {N. J. Phys.} \textbf{14}, 035019 (2012).
  
\end{thebibliography}

\begin{thebibliography}{10}
	
	\bibitem{Vasseur}
	R.~Vasseur, J.~P. Dahlhaus, and J.~E. Moore, 
	Phys. Rev. X \textbf{4}, 041007 (2014).
	
	\bibitem{BakGilPen09a} W.~S. Bakr, J.~I. Gillen, A.~Peng, S.~Folling, and M.~Greiner, 
	{Nature} \textbf{462}, 74 (2009).
	
	
\end{thebibliography}
\end{document}